\documentclass[12pt]{article}
\usepackage{epsfig}
\newcommand{\be}{\begin{eqnarray}}
\newcommand{\en}{\end{eqnarray}}
\begin{document}
\begin{center} {\bf\Large Resolution of a  long standing discrepancy in R 
with spin zero quarks}
\end{center}
\medskip
\centerline { S. Pacetti[1] and Y. Srivastava[1,2]}
{\small \medskip
\par\centerline {1. Dipartimento di Fisica \& INFN, Universit\`a di Perugia, 
Perugia, Italy}
\par\centerline {2. Physics Department, Northeastern University, Boston, 
MASS, USA}
$$ $$
\bigskip  
\centerline {\bf Abstract} 
A previously successful dispersive method has been applied to understand 
different values for $R(\sqrt{s}\ = 5 \div 7.5\ GeV)$ obtained by MARK I and 
Crystal Ball collaborations. We compute $R$ in the disputed region with 
data from outside this region and asymptotic behavior given by the standard 
model with 5 quark flavors, and find  agreement with the Crystal 
Ball result. On the other hand, the  MARK I data are reproduced
if we augment the asymptotic behavior with contributions from a single spin 
zero quark of charge ($ -1/3$). The visible hadronic fragments from
such scalar quarks are not likely to produce predominantly pure 2-jet 
events at such low energies. Hence, such decay modes may have been 
removed by the Crystal Ball energy imbalance cuts in their definition 
of hadronic events but not in MARK~I events, thus accounting  for the 
discrepancy in the two results. Upper bounds on spin zero quark production 
at LEP through $Z$ decay data are used to estimate the mixing angle 
between $T_3\ =\ -1/2$ and $T_3\ =\ 0$ scalar quarks. Recent negative 
results about spin zero quarks from CLEO are critically examined. We 
briefly discuss diquark production hypothesis and find it very unlikely 
to explain the discrepancy.    

$$ $$   
The purpose of this work is to shed some light on an ancient and as yet 
unresolved discrepancy in the measured values of $R(s)$ in the energy range 
$\sqrt{s}\ = 5\div 7.5\;GeV$, both measured at SLAC. The MARK I \& II 
collaborations [1,2] found a higher value $R_{av} = 4.3 \pm 0.4$, whereas 
the Crystal Ball collaboration [3] found a lower value 
$R_{av} = 3.44\ \pm 0.03\ \pm 0.18$. There are a few data points in this
region from PLUTO [4] and DASP [5] collaborations as well which are in
agreement with the MARK I collaboration. 
\par    We employ a dispersion-theoretic formalism which has been quite 
successful in studies of elastic form factors of nucleons and mesons 
[6-8]. For example, using space-like and above the $N\bar{N}$ threshold 
time-like data for the nucleon form factor, we were able to compute the 
same elastic nucleon form factor in the experimentally inaccessible region 
between the two pion and two nucleon thresholds, in a model independent way 
without any bias towards expected resonances. Remarkably,
resonance structures with peaks for the $\rho(770)$, $\rho'(1600)$ and a 
structure near the $N\bar{N}$ were automatically generated. For details 
about the regularization scheme employed to solve the integral equations, 
we refer the reader to references [6-8].\par

         For the problem at hand, we consider the (Lorentz scalar) vacuum 
polarization function $\Pi_\gamma(s)$ which is defined via the EM 
polarization tensor $\Pi_{\mu\nu}(s)$ as:
\be
ie^2\int d^4x e^{iqx}\langle 0|Tj^\mu_{em}(x)j^\nu_{em}(0)|0\rangle=
-(q^2g^{\mu\nu}-q^\mu q^\nu)\Pi_\gamma(q^2),\nonumber
\en
where $j^\mu_{em}(x)$ is the electro-magnetic current.
$\Pi_\gamma(s)$ is an analytic function in the complex $s$ plane with a 
branch cut 
on the real positive axis for $s>s_0$, where $s_0=4m_\pi^2$. We use the 
dispersion relation:
\be
\Pi_\gamma(t)-\Pi_\gamma(0)=\frac{t}{\pi}\int_{s_0}^\infty ds
\frac{Im \Pi_\gamma(s)}{s(s-t-i\epsilon)},
\en
subtracted at $t=0$, with the renormalized photon vacuum polarization:
$\Pi(t)=\Pi_\gamma(t)-\Pi_\gamma(0)$. The optical theorem relates 
the imaginary part of the vacuum polarization $\Pi(s)$ to the function 
$R(s)$ by means of the identity:
\be
Im\ \Pi(s)=\left( {{\alpha}\over{3}} \right) R(s).
\en
One obtains, therefore, the dispersion integral [9]:
\be
\Pi(t)={{t\alpha}\over{3\pi}}\int_{s_0}^{\infty}ds{{R(s)}\over{s(s-t)}},
\en
through which $\Pi$ for space-like $t$ gets related to $R$ for time-like
$s$.

 The idea is to consider this relation as an integral equation to
compute $R(s)$ in the interval $[5\div 7.5\ GeV]$, using the following 
input: (i) experimental data in the time-like region from threshold
up to $5\ GeV$ and from $7.5\div 100\ GeV$, (ii) PQCD  asymptotic behavior for 
the rest of the time like region and (iii) in the space-like region only 
the asymptotic value of $\Pi(t)$ calculated for values of $t$ in the
interval $[-205\ GeV^2\le \bar{t}\le-200\ GeV^2]$ through PQCD. 
Then we have:
\be
\Pi(t)-I_{01}(t)-I_{23}(t)-I_{a}(t)=
{{t\alpha}\over{3\pi}}\int_{s_1}^{s_2}ds{{R(s)}\over{s(s-t)}}
\en
where:
\be
I_{01}={{t\alpha}\over{3\pi}}\int_{s_0}^{s_1}ds{{R_{exp}(s)}\over{s(s-t)}},
\ \ \
I_{23}={{t\alpha}\over{3\pi}}\int_{s_2}^{s_3}ds{{R_{exp}(s)}\over{s(s-t)}},
\ \ \ 
I_{a}(t)={{t\alpha}\over{3\pi}}\int_{s_3}^{\infty}ds{{R_{a}(s)}\over{s(s-t)}},
\en
with: $s_1=(5\ GeV)^2$, $s_2=(7.5\ GeV)^2$ and $s_3=(100\ GeV)^2$.
The function $R_{exp}(s)$ used in the first and second integral (5) is a fit 
of the experimental data, the function $R_{a}(s)$ in the third integral 
is the PQCD asymptotic behavior at leading order:
\be 
R_a=N_c\sum_q Q_q^2\left[ 1+\frac{\alpha_s}{\pi} \right],
\en
where $N_c$ is the colour factor and $Q_q$ is the charge of the quark $q$.
We note that higher order QCD correction terms (that is, beyond the
$\alpha_s/\pi$ term) are indeed ``corrections to a correction'' and thus 
play a negligible role in our analysis of the discrepancy.\par 
The five light quark ($u,d,s,c,b$) contributions to $\Pi(t)$ can be safely
 calculated only at large $-t$, since at low energies the quark interactions 
are  modified considerably by strong interactions. 
At high energies, by virtue of asymptotic freedom inherent in QCD, we can 
treat the quark contribution similar to that for the leptons and, at 
leading order, we obtain:
\be
\Pi(t)=-N_c{{\alpha}\over{3\pi}}\sum_{q}Q^2_q\left[ \ln\left(-
{{t}\over{m_q^2}} \right)-{{5}\over{3}}+O\left(-{{m_q^2}\over{t}} 
\right) \right].
\en
The difficulty in using this function comes from the masses, which is
particularly severe for the light quarks $u$, $d$ and $s$, that are not 
unambiguously defined. To avoid this problem, but without any further 
approximation, we consider the derivative in $t$ of the dispersion integral 
(3) so as to obtain a new integral equation:
\be
{{d\Pi(t)}\over{dt}}-{{dI_{01}(t)}\over{dt}}-{{dI_{}(t)}\over{dt}}
-{{dI_a(t)}\over{dt}}=
{{\alpha}\over{3\pi}}\int_{s_1}^{s_2}ds{{R(s)}\over{(s-t)^2}}. 
\en
The asymptotic behavior of ${{d\Pi(t)}\over{dt}}$ for large
$-t$ is:
\be
\left[ {{d\Pi(t)}\over{dt}} \right]_A = -N_c{{\alpha}\over{3\pi}}\sum_{q}
Q^2_q{{1}\over{t}}.
\en
It has not just the virtue of being independent of the quark masses: 
it requires only a knowledge of the quark charges and the colour factor. 
Upon solving the integral equation (8) [6-8], we can calculate the values of 
$R(s)$ in the disputed region $[s_1,s_2]$ using as input the time-like 
experimental data and the quark charges.

 Including only the five standard (spin 1/2) light quarks $u,d,s,c,b$, 
as shown in Fig.(1a), we find a value for $R(s)$ in good agreement with 
the Crystal Ball data [10].\vspace{-.5cm} 
\begin{figure}[h]
\begin{center}
\epsfig{file=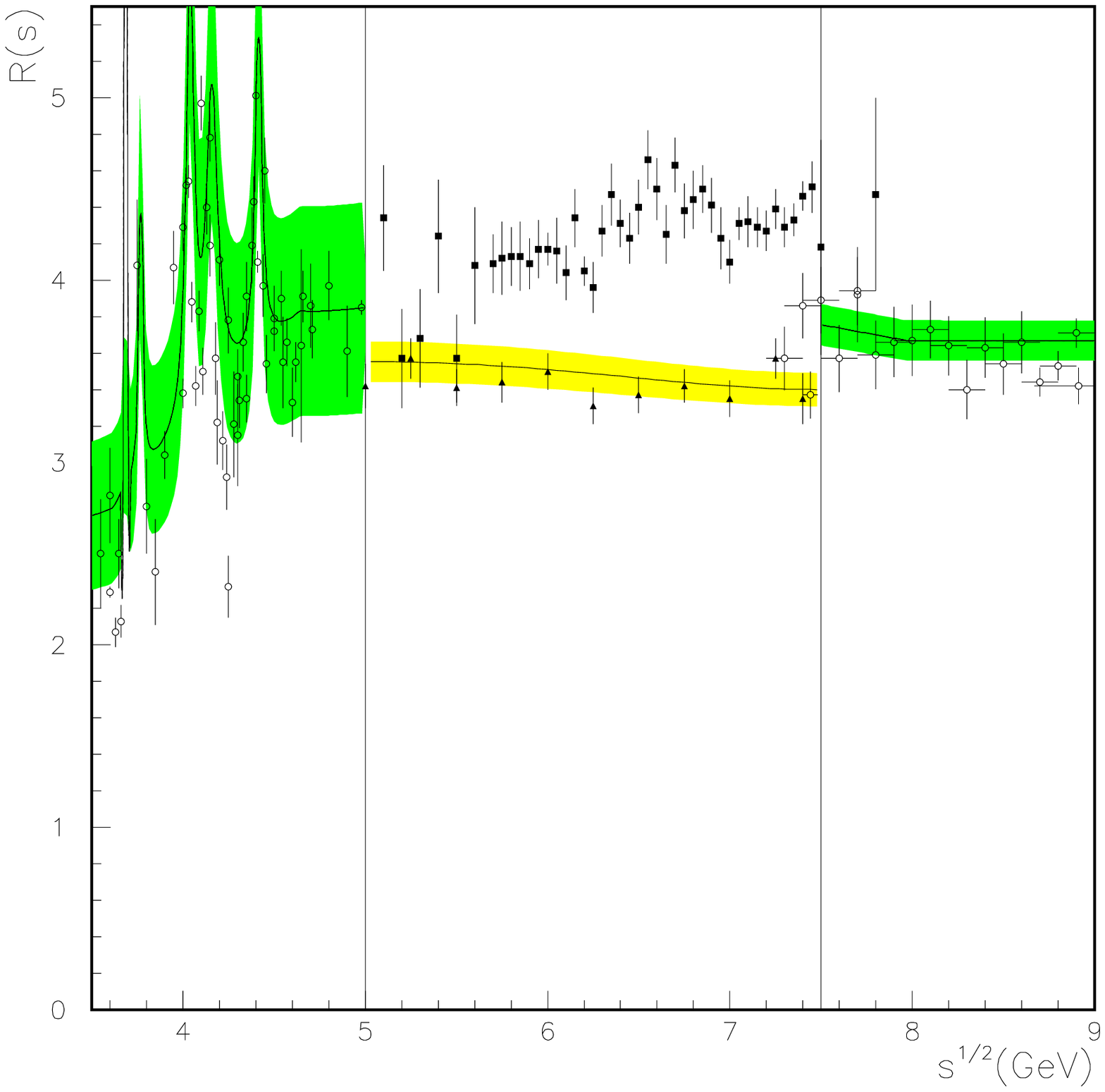,height=9cm,width=9cm}\\
\begin{minipage}{9cm}\small
{\bf Figure 1a.} Values of R(s) (light grey band) obtained from 
integral equation (8) with five spin 1/2 quarks in the interval 
$[5GeV,7.5GeV]$ of $\sqrt{s}$.
\end{minipage}
\end{center}
\end{figure}\vspace{-.5cm}\\ 
But, if we include an additional contribution 
from a single species of spin zero quark of charge (-1/3), which requires 
simply adding in the summations (6) and (9) another charge (-1/3) squared, 
multiplied by a factor 1/4 (due to zero spin of the quark), we reproduce 
the MARK I data, shown in Fig.(1b).
\newpage 
\begin{figure}[t]
\begin{center}
\epsfig{file=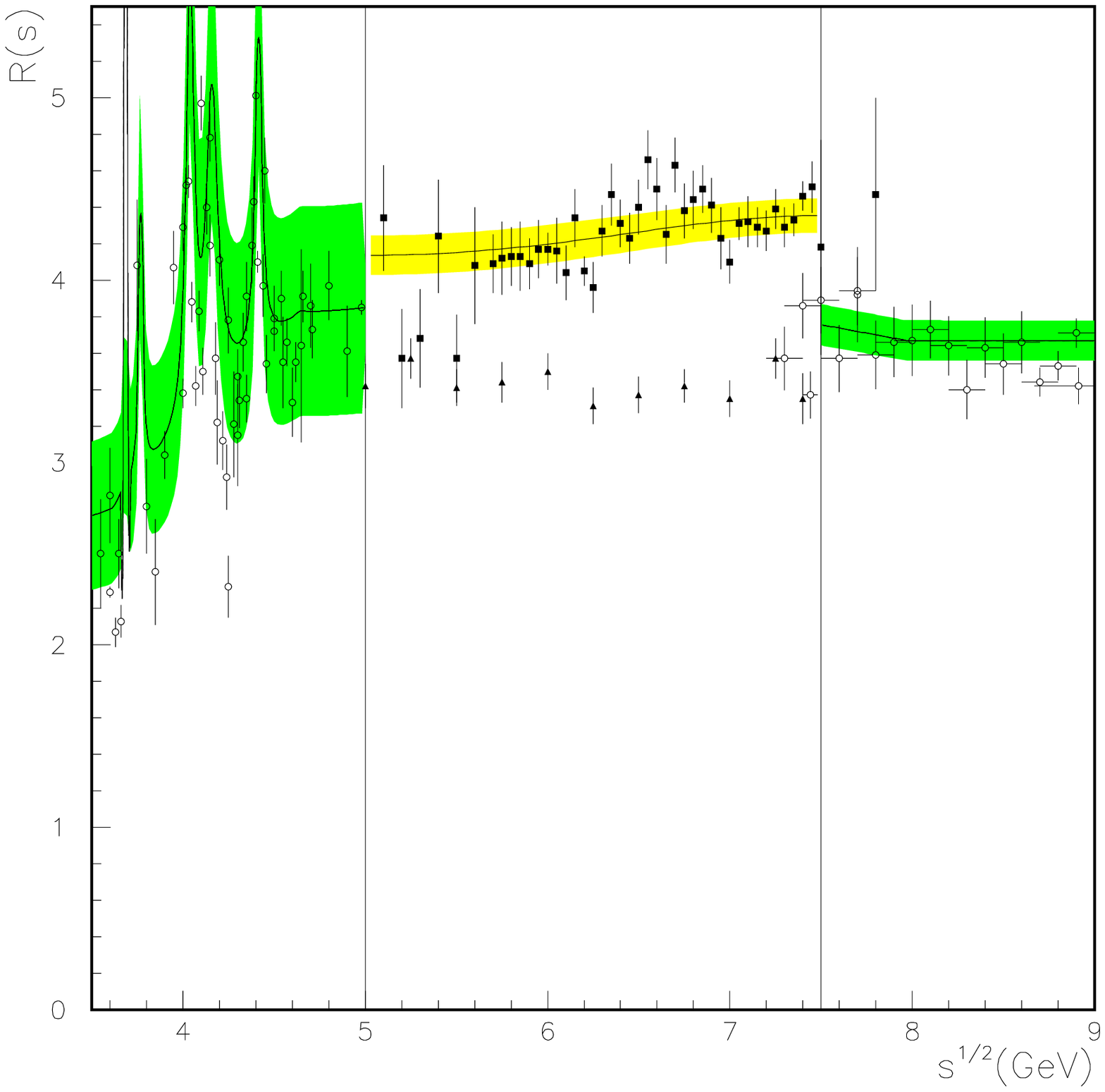,height=9cm,width=9cm}\vspace{-.3cm}\\
\begin{minipage}{9cm}\small
{\bf Figure 1b.} Values of R(s) (light grey band) 
obtained from integral equation (8) with
five spin 1/2 quarks and one scalar quark.
\end{minipage}\vspace{-1cm}
\end{center}
\end{figure}
\noindent 
In Fig.(2a) and Fig.(2b), we show a similar comparison but imposing continuity 
at the boundaries\vspace{-.5cm}.
\begin{figure}[h]
\begin{center}
\epsfig{file=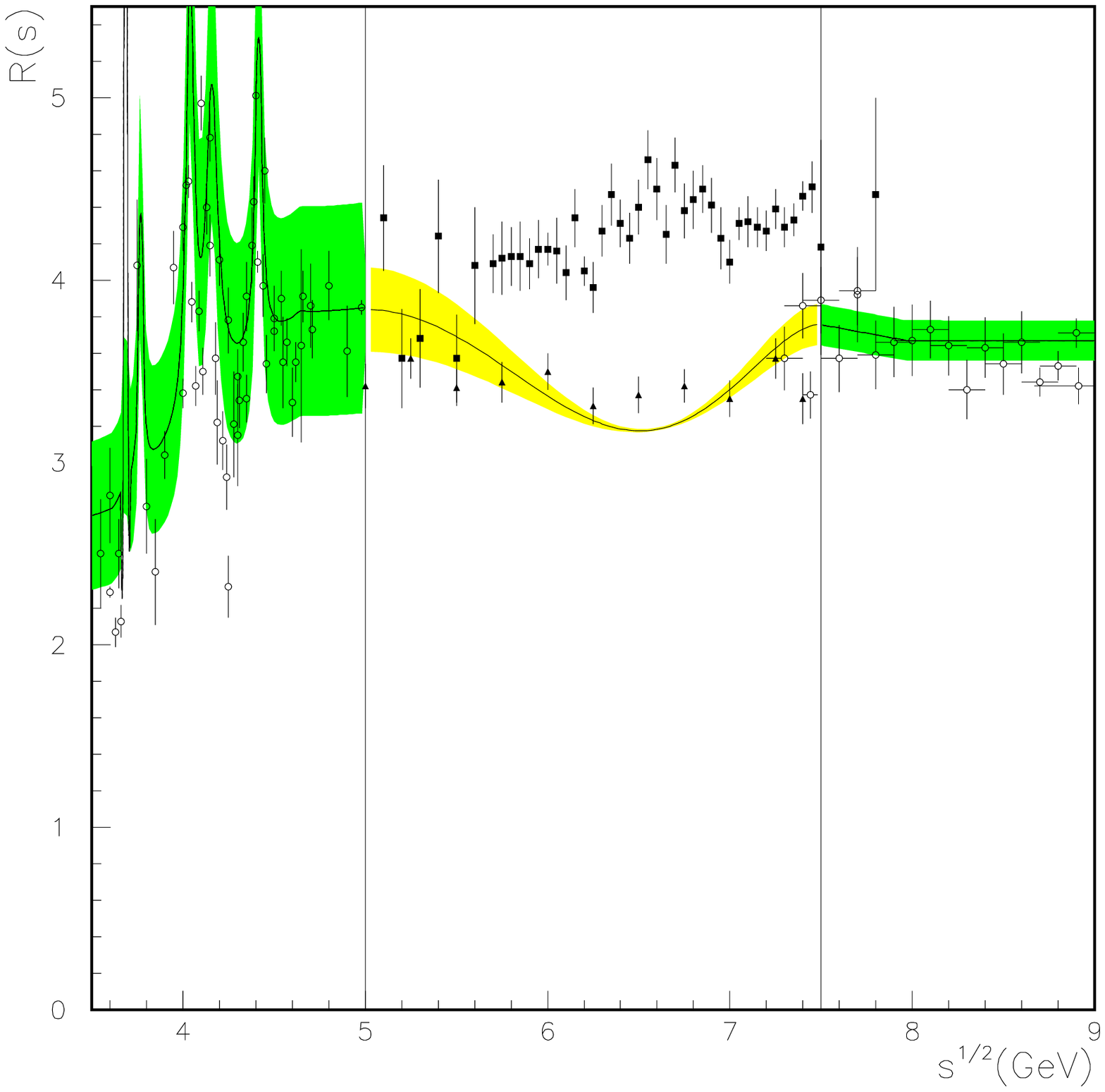,height=5.5cm,width=7cm}\hspace{1cm}
\epsfig{file=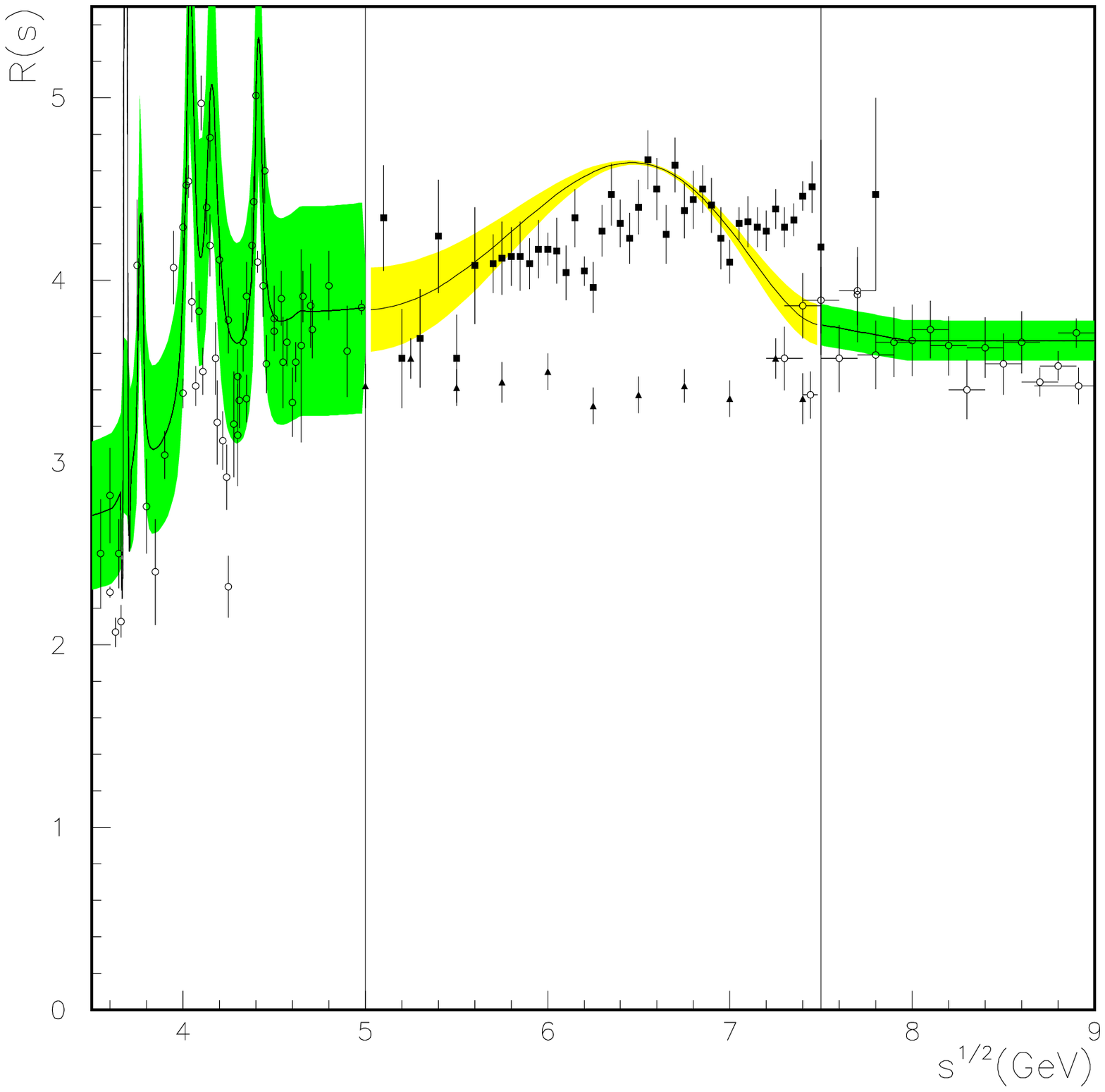,height=5.5cm,width=7cm}\vspace{-.2cm}
\begin{minipage}{7cm}\small
{\bf Figure 2a.} Values of R(s) (light grey band) obtained from 
eq. (8) with five spin 1/2 quarks. Continuity imposed at 5 GeV
and 7.5 GeV.
\end{minipage}\hspace{1cm}
\begin{minipage}{7cm}\small
{\bf Figure 2b.} Values of R(s) (light grey band) obtained from 
eq. (8) with five spin 1/2 quarks and one scalar quark. Continuity
imposed at 5 GeV and 7.5 GeV.
\end{minipage}
\vspace{-.8cm}
\end{center}
\end{figure}\\
        As a consistency check, we write a dispersion relation for 
$\Delta R(s)$ in the above region whose asymptotic value contains
contribution from one species of colour triplet, charge $(-1/3)$, 
spin zero quark only. Phenomenologically, in the range 
$s_3\ \ge\ s\ \ge\ s_2$, the standard 5 quark model gives a rather 
good description of the data. Hence, we obtain the approximate relation
\be
\int_{s_1}^{s_2} \frac{\Delta R(s)ds}{(s - \bar{t})^2}
\approx  - \frac{1}{12 \bar{t}}\left[1 + \frac{\bar{t}}{(s_3 - \bar{t})}
\right]
\en
This gives an average value $\overline{\Delta R}\ \approx 0.75$. By way
of comparison, the difference $\Delta R(s)$ between the two curves 
from Fig.(1) are plotted in Fig.(3a). Fig.(3b) shows the difference 
$\Delta R(s)$ for \vspace{-.5cm}Fig.(2). 
\begin{figure}[h]
\begin{center}
\epsfig{file=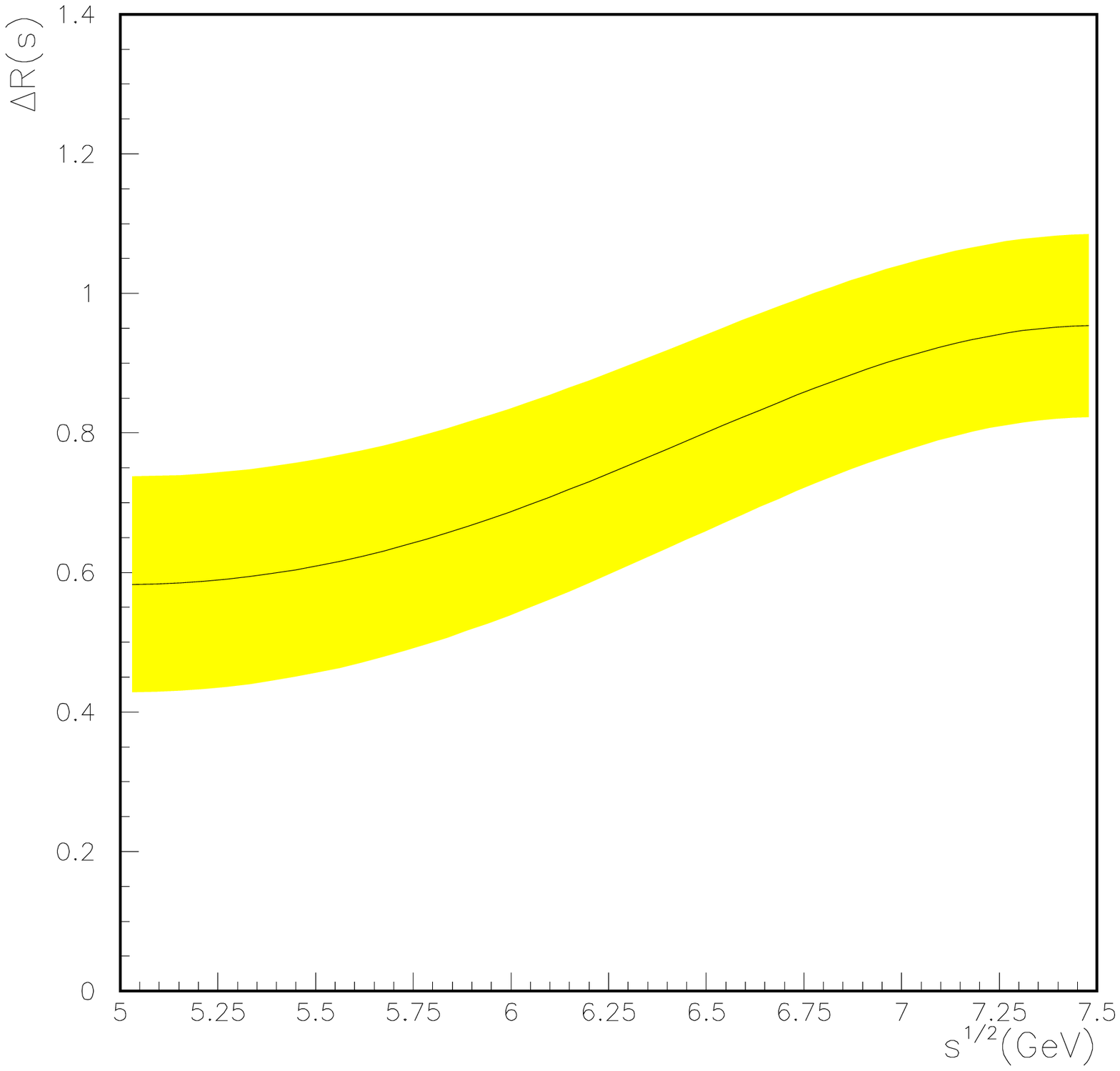,height=6.cm,width=7cm}\hspace{1cm}
\epsfig{file=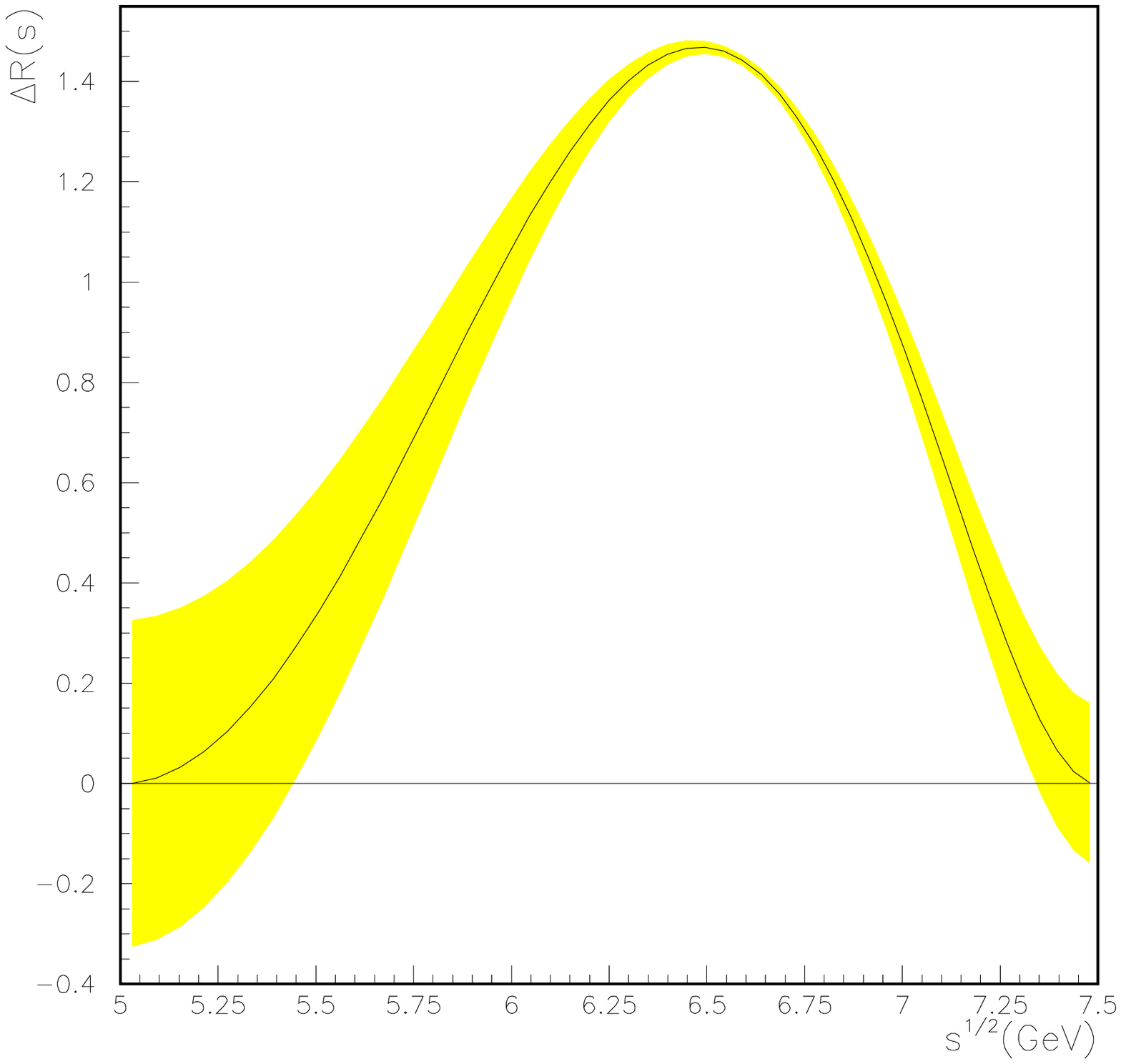,height=6.cm,width=7cm}\vspace{-.3cm}
\begin{minipage}{7cm}\small
{\bf Figure 3a.} $\Delta R(s)$ from Fig.(1).
\end{minipage}\hspace{1cm}
\begin{minipage}{7cm}\small
{\bf Figure 3b.} $\Delta R(s)$ from Fig.(2).
\end{minipage}
\vspace{-.7cm}
\end{center}
\end{figure}\\
These average values are mutually consistent. Such a ``locally averaged'' 
$\overline{\Delta R}$ should not be confused with $\Delta R_{asymptotic}$. 
The above has been obtained under the assumption that beyond the disputed
region, experimental data are adequately described by the standard 5 
spin $1/2$ quark model. Thus, roughly $(\Delta R_{asy}\times (-\bar{t}) 
\approx\ (\overline{\Delta R})\times (s_2 - s_1))$. 

        The two experiments can be reconciled only under the hypothesis
that Crystal Ball experimental cuts may have eliminated signals due to the 
spin zero quarks. The bound states and resonances with $J^{PC}\ =\ 1^{--}$
coupling to $e^+e^-$ are in relative $P$- wave for spin zero quarks. They
would be relatively close in energy and more importantly, are expected to 
decay copiously into a low energy photon ($E_\gamma\ \approx 300\ MeV$) 
recoiling against $S$- wave states which would subsequently decay into 
hadrons via 2 gluons [11]. Such radiative decay events would have a very 
asymmetric energy pattern. For the model discussed below, a similar 
pattern may also follow (for the non resonant) scalar quark production.
In their definition of hadronic events contributing to $R$, Crystal Ball
imposed the following kinematic cuts for the energy imbalance between
left-right, top-bottom and front-back hemispheres. If any of these fractional
energy differences, $A_{left-right}$, $A_{top-bottom}$ or $A_{front-back}$
was less than $0.8$, the event was classified as being due to beam-gas
interactions, $\gamma\gamma$ collisions or large missing energy $\tau$ 
decays. Thus, such events were not included in $R$ by Crystal\vspace{-.7cm} 
Ball.
\begin{figure}[h]
\begin{center}
\epsfig{file=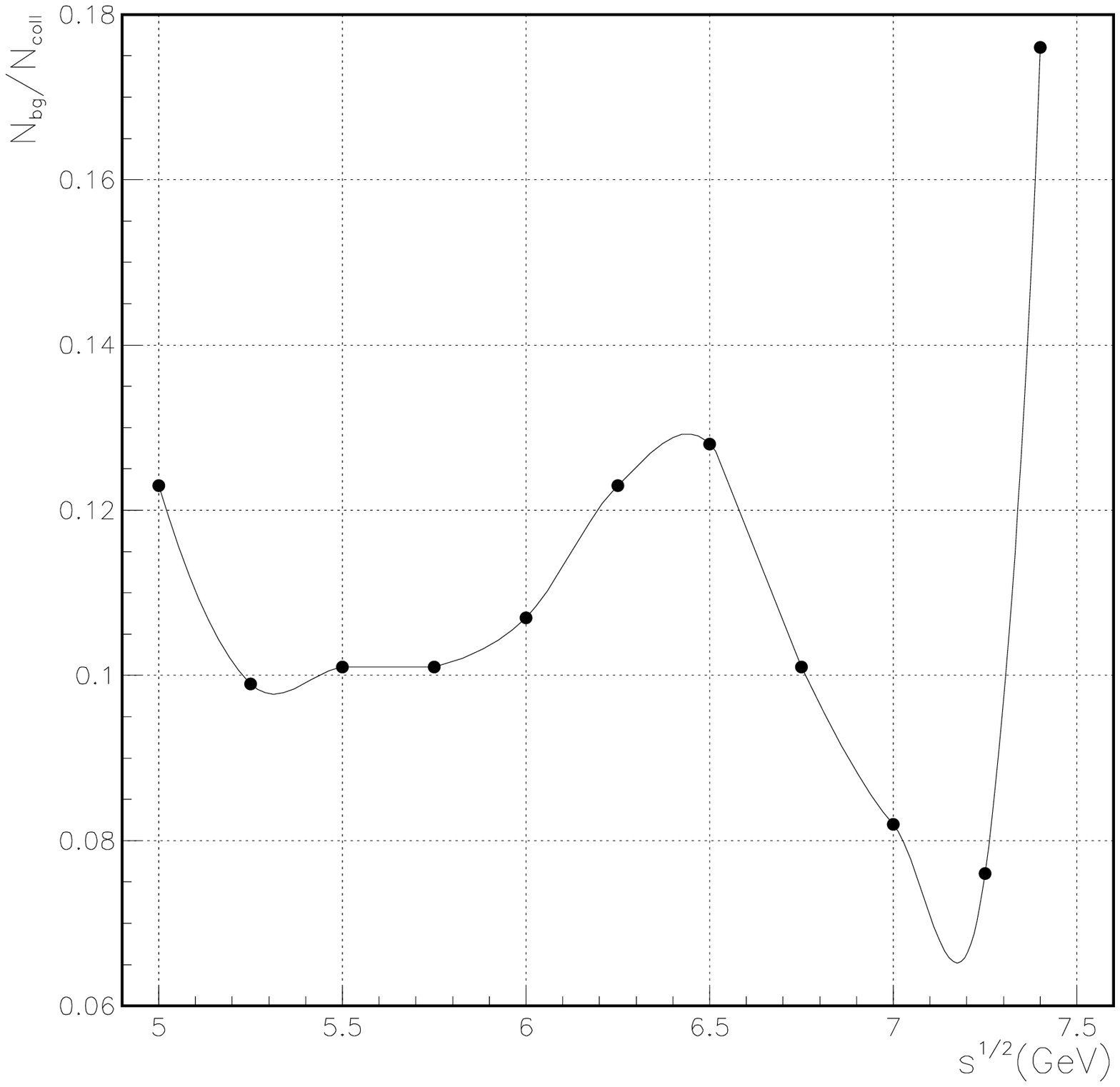,height=5.7cm,width=8cm}\vspace{-.2cm}
\begin{minipage}{8cm}\small
{\bf Figure 4.} Percentage of background events discarded by Crystal Ball.
\end{minipage}
\end{center}\vspace{-.3cm}
\end{figure}\\
In Fig.(4), we show a plot of the ratio between $N_{bg}$, which is the
number of the background events from beam-gas and beam-wall interactions,
and the total collisions $N_{coll}$ passing the hadron selection criteria,
versus $\sqrt{s}$ from Table II of ref.(3). 
That the rejected events have peaks and structures resembling the MARK I 
data appears significant and supports our hypothesis.

Now we turn to a rough estimate of the mass of such a scalar quark (Y). 
Taking our cue from the steep rise in the CB rejected events between 
$7.2\div 7.4\ GeV$, we may estimate $m_Y\ \approx\ 3.6\div 3.7\ GeV$ from the
production threshold for a pair ($Y^\dagger Y$) of such quarks. The local rise 
in MARK I data, prior to this threshold, beginning around $5\ GeV$ would 
however 
require that a single scalar quark be produced along with a couple of other 
light (spin $1/2$ $u,d,s,c$) quarks. One possibility (out of many others) 
would be a charge $(-1/3)$ color triplet scalar quark carrying baryon number 
$(-2/3)$ to coincide with the quantum numbers of a standard $\bar{U} \bar{D}$ 
(where $U\ =\ u,c$ and $D\ =\ d,s$) quark pair. Then, the final state 
$YUD$ would be allowed in $e^+e^-$ reaction, with a threshold 
around $5\ GeV$. We expect no sharp structures associated with it and
the level of production should decrease rapidly with $s$ so as to reach
its asymptotic parton level. Two points are worthy of note here. First, 
such a scalar quark would be an ``elementary'' particle and not a 
composite ``diquark'' state of two standard $\bar{U}$ and $\bar{D}$ 
quarks. (For this case, there might be non trivial mixing and interference
terms). While for the narrow window of pair produced resonances 
($Y^\dagger Y$ states), non-relativistic potential models may be a reasonable 
rough guide, the production dynamics and level of cross section for 
multi-particle relativistic systems (with exotic quantum numbers such as 
$YUD$) are not easy to compute or estimate directly since non relativistic 
potential models can not be employed for this purpose. The hadronization of 
scalar quarks would also be very different from those for the other quarks. 
Not knowing the nature of the beast, our modest aim here has been to 
estimate these contributions to R through unitarity, dispersion relations 
and asymptotic behavior which are consistent with data from outside 
this region. Secondly, if one were to study the mass distribution of an 
$e^+e^-$ or $\mu^+ \mu^-$ in the final state produced say from a 
$p\bar{p}$ initial state, we would expect to see an enhancement visible 
only around $7.2\div 7.4\ GeV$ (due to pair production of scalar quarks) 
since the probability to find $YUD$ in the initial state would be 
negligible.   

An alternative explanation of this rather substantial difference
between MARK I and CB data has been offered by Eidelman and Jegerlehner [12].
According to them, this difference arises from QED and QCD radiative
corrections. In the final analysis, our explanation of this difference
also involves radiative events but its source is different. Ours is
generated from an extra scalar quark and MARK I data are ``physical'',
whereas in the other explanation MARK I data should be divided 
(``corrected'') by the radiative effects. It should be firmly kept in
mind that the ($g\ -\ 2$) results discussed in ref.(12) can not be used to
discriminate between MARK I and CB data since the results are hardly 
changed whether one or the other data are used. 

The non resonant part of events from scalar quark decays would also
not be of the 2-jet type at low energies. Both in the ALEPH [13] and the
CLEO [14] collaboration searches for scalar quarks,  it has been assumed that 
each scalar quark decays weakly, viz., into $c\mu^-\tilde{\nu}_s$
and  $ce^-\tilde{\nu}_s$, where $\tilde{\nu}_s$ is a scalar neutrino.
For $\sqrt{s}=5\div 7.5\ GeV$ and scalar quark mass in the $3\div 4\ 
GeV$ range, the decay events would not be of the 2-jet type. On the
other hand, at much higher energies, for example $\sqrt{s}=160\div 
205\ GeV$ as in the ALEPH data, the decay products would be confined to 
the forward and backward directions similar to a 2-jet profile.

Evidence supporting a low mass spin zero quark must now be confronted with 
the extremely accurate decay systematics for the $Z^o$ for various channels 
available from LEP [15-18]. A computation of the coupling of observable 
spin zero quarks to the $Z$ boson requires a knowledge of their weak 
iso-spin. Let $Y_L$ and $Y_R$ denote spin zero quarks with 
$T_3\ =\ -1/2$ and $T_3\ =\ 0$ respectively. After mixing, let $Y_-$ and 
$Y_+$ denote the low and high mass eigenstates. If the higher of these 
masses happens to be larger than the $Z^o$ mass, kinematically $Z^o$ can decay 
only into $Y_-^\dagger Y_-$, with the decay amplitude given by
\be
Amplitude (Z^o(P)\rightarrow Y_-^\dagger(p_1) Y_-(p_2))
= \left( {{e}\over{sin\vartheta_W cos\vartheta_W}} \right) 
K \epsilon_\mu(P) (p_1 - p_2)^\mu , 
\en
where $K$ is given by
\be
K = \left( {{1}\over{3}}sin^2\vartheta_W \right) 
|U_{R-}|^2
+ \left( -{{1}\over{2}} + {{1}\over{3}}sin^2\vartheta_W \right) |U_{L-}|^2.
\en
Here $U_{R-}\ =\ cos\delta$ and $U_{L-} =\ sin\delta$ are 
the mixing matrix elements to the lower mass state ($-$). Thus, the 
branching ratio of $Z^o$ into a pair of low mass spin zero quarks can be 
arbitrarily small, even zero for $\delta^*$ 
\be
\delta^* = arcsin\left( \sqrt{{2sin^2\vartheta_W\over{3}}}
\right) \approx 23^o.
\en   
Present measurements of the $Z^o$ width and observed branching 
ratios therefore can serve to place limits on the mixing angle $\delta$. 
For example, if we assume that the branching ratio for the spin zero quark 
decay channel is less than one part per mille, approximately
$\delta\ =\ (23\ \pm\ 15)^o$ (for $\delta$ chosen in the first quadrant). 

The recent CLEO experiment has given quite stringent bounds regarding
a low mass squark partner of the $b$ quark. They have looked for and 
not found an expected high level $D$ and $D^*$ signal. Of course, it
could be that the low mass spin zero quark is not the partner of
the $b$ quark or even if it were so, the squark mixing matrix may not 
be similar to the quark mixing matrix. Also, the estimate of the weak
decay signal expected in the CLEO experiment is based on phase space, i.e.,
a structureless matrix element and an essentially zero mass scalar 
neutrino. Neither of these hypotheses can be justified theoretically.
For example, even a $300~MeV$ neutrino would lower the signal to practically
its background value. We are not aware of any argument which can so
restrict the scalar neutrino mass.   

        Among other presently available beams, high luminosity asymmetric
B-factories offer a good avenue for new and independent check of
 data on $R$ in the $5\div 8\ GeV$ region, by observing a bremsstrahlung 
photon of energy $1.3\div 3.8\ GeV$. An independent check of scalar quark
production at B-factories would be through their weak decays by looking
at changes in the number of opposite sign leptons (in the opposite 
hemispheres) as the beam energy passes through the $b\bar{b}$ threshold. 
Signals from spin zero quarks should be looked for also in $\mu^+\mu^-$ 
production and in the hadronic machines via the gluon induced process 
$g+\bar{D}\rightarrow Y+U$ (for the mechanism discussed earlier). 
In addition, the next Tevatron run offers the possibility of looking for 
$Y$ jets as well as a means to  probe for $\tilde{w}$ (any generic spin 
$1/2$ object recoiling against the $Y$) up-to a mass of about $170\ GeV$ 
through the top quark decay  $t\rightarrow\ Y\ +\ \tilde{w}^+$.
 
We now discuss a completely different hypothesis to resolve the 
discrepancy. If one accepts that the difference between the two 
experiments is due to one experiment being essentially sensitive only
to 2-jet events and the other not so biased, this may be accounted for
through the production of a diquark-antidiquark pair, since they
would produce predominantly 4 quark final states. As the diquark photon
vertex includes a form factor their production would disappear at
high energies, e.g., at LEP.  

A diquark diagram (inclusive of a form factor) to the self energy 
of the photon is shown in Fig.(5). 
\begin{figure}[h]
\begin{center}
\epsfig{file=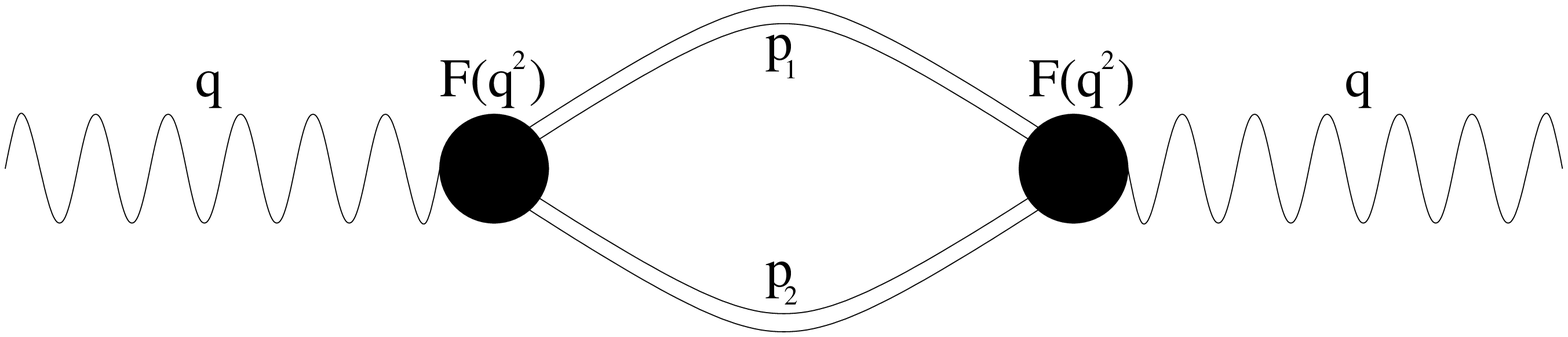,height=2cm,width=9.cm}\vspace{.3cm}\\
\begin{minipage}{9cm}\small
{\bf Figure 5.} Diquarks photon self energy.
\end{minipage}
\end{center}
\end{figure}\vspace{-.5cm}\\
It contributes to the imaginary part
\be
\mbox{Im}(\pi_{DQ}(s))=\frac{1}{4}\frac{N_c\alpha Q^2}{3}|F(s)|^2
\left(1-\frac{4m^2}{s}\right)^\frac{3}{2}
\en
and thus to the dispersion relation
\be
\hat{\pi}_{DQ}(t)=\frac{1}{4}\frac{N_c\alpha Q^2}{3}\frac{t}{\pi}\int_{4m^2}
^\infty\left(1-\frac{4m^2}{s}\right)^\frac{3}{2}\frac{|F(s)|^2}{s(s-t)}ds.
\en
$F(s)$ is normalized to $1$ at $s\ =\ 0$ and is expected to go asymptotically 
as
\be
|F(s)|\sim\frac{Q_0^2}{s},
\en
with $Q_0^2\simeq 10\;GeV^2$ [19]. For a diquark of charge $q$ to 
saturate the discrepancy in $\overline{\Delta R}=0.75$ (as discussed 
previously), we require 
\be
|F(s)|=\left\{\begin{array}{ll}
\frac{3}{N_cq} & \mbox{if }s\in[s_1,s_2]\\
N_cq\frac{Q_0^2}{s} & \mbox{if }s>s_2\\
\end{array} \right.,
\en  
to be fed in
\be
\frac{d\hat{\pi}_{DQ}(t)}{dt}=
\frac{\alpha}{3\pi}\frac{1}{12}
\int_{s_1}^\infty\left(1-\frac{s_1}{s}\right)^\frac{3}{2}
\frac{|F(s)|^2}{(s-t)^2}ds.
\en
Unfortunately, the contribution of such diquarks in the space like 
region falls short by a factor $4$. Hence, the hypothesis of diquarks 
being responsible for the discrepancy does not appear to be internally
self consistent. 

        In conclusion, our dispersive analysis of the MARK I data
(the {\it only set} included in PDG, the particle data tables [20]) 
appears to require the existence of a charge $(-1/3)$ scalar quark with 
a low mass. While the level of the reported cross section 
appears to be consistent with our analysis, it remains an open dynamical
problem how to generate such a large value. Preliminary estimates of 
contributions to $R$ from bound states computed through static potentials 
appear to fall short by a factor of 3 or more. This problem obviously
deserves further study and the expected data from BABAR on a new measurement
of $R$ in the disputed region suggested here would be most useful. On the  
 other hand, lack of  evidence in the present LEP data about 
the direct production of such low mass scalar quarks in $Z^o$ decay only 
serves to indicate that it is predominantly an iso-singlet. On the other 
hand, a $b$ jet from the $Z^o$ decay may sequentially produce such a scalar 
quark whose signature might be revealed through an analysis of the final 
$\mu^+\mu^-$ mass distribution in the $7.2\div 7.4\ GeV$ region.   

        It is with much gratitude that we thank P. Giromini for directing
our attention to this problem and for fruitful discussions at every stage 
of this work. It is a pleasure to thank S.~Glashow, U. Heintz, K. Lane, 
M. Narain, G. Pancheri, S. Reucroft, J. Swain and A. Widom for much 
encouragement and many helpful suggestions.

\newpage
\centerline {\large{\bf Footnotes and References}}            
\vspace{.2cm}  
\begin{enumerate}
\item J. L. Siegrist {\it et al.} MARK I Coll., {\it Phys. Rev.} 
{\bf D26} (1982) 969.
\item M. W. Coles {\it et al.} MARK II Coll., {\it Phys. Rev.} 
{\bf D26} (1982) 2190.
\item C. Edwards {\it et al.} Crystal Ball Coll., SLAC-PUB-5160 
(1990).
\item J. Burmeister {\it et al.} PLUTO Coll., {\it Phys. Lett.}
{\bf 66B} (1977) 395; Ch. Berger {\it et al.}, {\it Phys. Lett.}
{\bf 81B} (1979) 410.
\item R. Brandelik {\it et al.}, DASP Coll. {it Phys. Lett.}
{\bf 76B} (1978) 361; H. Albrecht {\it et al}, {\it Phys. Lett.}
{\bf 116B} (1982) 383.
\item R. Baldini, S. Dubni\v{c}ka, P. Gauzzi, E. Pasqualucci, S. Pacetti
and Y. Srivastava, {\it Euro. Phys. J.} {\bf C11} (1999) 709.
\item R. Baldini, S. Dubni\v{c}ka, P. Gauzzi, E. Pasqualucci, S. Pacetti
and Y. Srivastava, {\it Nuc. Phys.} {\bf A666\&667} (2000) 38c.
\item R. Baldini, S. Dubni\v{c}ka, P. Gauzzi, E. Pasqualucci, S. Pacetti
and Y. Srivastava, Frascati Report LNF-98/024(IR).
\item N. Cabbibo and R. Gatto, {\it Phys. Rev. Lett.} {\bf 4} (1960) 313;
{\it Phys. Rev.} {\bf 124} (1961) 1577.
\item R. M. Barnett, M. Dine and L. McLerran, {\it Phys. Rev.} {\bf D22}
(1980) 594. 
\item C. R. Nappi {\it Phys. Rev.} {\bf D25} (1982) 84.
\item S. Eidelman and F. Jegerlehner, {\it Z. Phys.} {\bf C67} (1995) 585. 
\item G. Taylor, LEPC Presentation, 20 July 2000, http://alephwww.cern.ch. 
\item V. Savinov {\it et al.} CLEO Coll., hep-ex/0010047. 
\item P. Abreu {\it et al.} DELPHI Coll. {\it Nucl. Phys.} {\bf B418} (1994) 
403.
\item M. Acciarri {\it et al.} L3 Coll. {\it Z. Phys.} {\bf C62} (1994) 551.
\item R. Akers {\it et al.} OPAL Coll. {\it Z. Phys.} {\bf C61} (1994) 199.
\item D. Buskulic {\it et al.} ALEPH Coll. {\it Z. Phys.} {\bf C62} (1994) 539.
\item M. Anselmino, E. Predazzi, S. Ekelin, S. Fredriksson and D.B. Lichtenberg, 
         {\it Rev. Mod. Phys.} {\bf 65} (1993) 1199.  
\item Particle Data Group, {\it Review of Particle Physics}, 
{\it Eur. Phys. J.} {\bf C3} (1998) 227. 
\end{enumerate}
\end{document}